\documentclass[11pt]{article}
\usepackage[utf8]{inputenc}

\newtheorem{defi}{Definition}
\newtheorem{teo}{Theorem}

\newtheorem{exi}{Example}
\newtheorem{propi}{Proposition}

\newtheorem{lemi}{Lemma}

\usepackage[sc]{mathpazo}
\usepackage[T1]{fontenc}

\begin{document}
\title{Local and Global Optima: Sheaves and Polynomial Approximations}
\author{Fernando Tohm\'e\\
Departamento de Econom\'{\i}a - Universidad Nacional del Sur\\
Instituto de Matem\'atica de Bah\'{\i}a Blanca - CONICET\\
{\tt ftohme@criba.edu.ar}}
\date{}
\maketitle
\begin{abstract}
In this paper, we analyze how global optima of an agent's preferences can be reconstructed from the solutions
found for local problems. A sheaf-theoretic analysis provides an abstract characterization of the global solution, and polynomial approximations are obtained when only a few local instances are available.
\end{abstract}

\section{Introduction}
An appropriate description of the behavior of economic agents is {\it as if} they run algorithms that translate given external states of the world into specific actions. Most of the economic literature roughly assumes that for each agent: $(i)$ there exists a real-valued function $U$ over outcomes ranking the well-being attributed by the agent to each of them; $(ii)$ that the agent has the subjectively perceived goal of achieving the maximum of $U$; $(iii)$ the agent consistently and systematically selects the maximum of $U$. In stronger versions the agent is assumed, $(iv)$ to be able to specify the set of feasible alternatives, given by the objective conditions in which he has to make a decision. Furthermore, $(v)$ the individual is able to find the actions that maximize $U$ under those constraints.

From a logical point of view, these propositions can be considered separately. Different specifications of the decision processes can result from considering combinations of yes-no answers to $(i), \ldots, (v)$, from the strictly defined {\em rational economic agent} ({\bf REA}) to what we could be called a {\em ``Dostoievskian'' agent}.\footnote{ This notion is due to Ana Marostica and Daniel Heymann. The Underground Man in Fyodor Dostoyevsky's {\em Notes from Underground}, avoids choosing the outcome that maximizes $U$, just to show that he is free and can overcome the determinism implicit in that choice.} We will focus on a particular instance, in which the agent tends towards the maximum of $U$, by solving constrained versions of the choice problem and using the solutions to build up closer approximations to $U$.

As a point of departure for our analysis, notice that the maximum of $U$ was obtained as a solution to a global optimization problem. But the practicing economist (or the theorist, when building a specific model) represents the agent in a {\em partial} setting, where it is assumed that the solution does not result as a constrained version of a pre-defined global solution. Furthermore, the solution to the partial problem is seen as independent from the choices the agent could make in other contexts.\footnote{This is the realm of Partial Equilibrium Analysis, which involves ``demand'' and ``supply'' functions of certain goods of interest. See \cite{Mas-Colell}.}

It is rather easy to think of conditions under which several partial problems should be solved together, without allowing a decomposition in independent parts. The reason why this is not the usual case in the economic literature is straightforward, at least from a heuristic point of view: trying to represent and solve the global maximization problem of an agent is likely to result in an intractable problem, and the gains in analytical power may not even be large.

The standard practice of describing {\em model-specific} optimizations could be associated with a pragmatic approach, quite different from maximizing $U$. This procedure starts with the identification (by the agent) of a {\em problem}. Then, a problem-specific function $u$ arises, focused on the arguments relevant to the question at hand. This $u$ is maximized, yielding the problem-specific solution.

The model-specific scenario leads to a broader approach to agent behavior. Here the agent is not cognizant of his whole set of choices but acts by identifying specific problems and solving them one by one. The solutions of the specific problems may remain as independent pieces of information. In this paper we will focus on how they may lead to a learning process, in which the agent gathers information, first solving particular problems and from then {\em abducting}\footnote{See \cite{Gabbay} for a thorough discussion of the scope of this inference mechanism.} hypotheses about the shape of the unknown $U$. This approximation may yield a function $V$, whose maximum can be contrasted, in a new problem with the solution of the maximization of the specific $u$. This allows to correct the specification of $V$. Repeated instances of this procedure yield, in the limit, a $V$ that can be identified with $U$.

It is easy to think of stories supporting this approach. Individuals living in a steady environment tend to choose time-revered solutions, but once their world is unsettled by, say, an external shock, they will try to extrapolate them to the new situation. But only those agents that learn how to adapt this generalization to the new circumstances will be able to survive.

An evolutionary approach is not the focus of this paper. Instead, we will seek to characterize the process by which an agent may approximate the solution of the {\em global} optimization problem up from a set of {\em local} solutions (each for a different problem). A simple criterion of minimality, very much like {\em Occam's Razor} will help to keep the simplest hypotheses. While the informal description is intuitively clear, the features that ensure a good approximation are rather involved. In technical terms, we are concerned with a {\em sheaf construction} under which a global solution under $U$ is obtained from the solutions corresponding to a class of model-specific $u$ functions. We will be concerned, in particular, with polynomial approximations, which can be patched up with relative ease.

The rest of this paper is organized as follows. Section 2 introduces a formal description of the problem. Section 3 provides a sheaf construction that allows to obtain global solutions up from a family of local utility functions. Section 4 considers the particular case of concave utility functions while section 5 presents the general case, analyzed through polynomial approximations. Section 6 discusses the evolution of global solutions. Section 7 concludes.

\section{Decision-making: local vs. global}
The traditional characterization of decision-making by an individual is as follows. Let ${\mathcal L}_i$ be a space of possible {\bf alternatives} that an agent $i$ may select.\footnote{The meaning of these alternatives depends on the context. If the agent is a consumer in a competitive market with a finite number of goods, she has to choose a vector of those commodities. In a planning problem, she has to select a plan specifying the amounts of resources used or consumed at each period, etc.} Each $x \in {\mathcal L}_i$ is evaluated by means of a {\em utility} function, $U_i: {\mathcal L}_i \rightarrow \Re$. Given a family of constraints that limit the set of options open to the agent to $\hat{L} \subseteq {\mathcal L}_i$, the goal of the agent is to find a $\hat{{\mathbf x}}$ that maximizes $U_i$ over $\hat{L}$.

To proceed, we fix the agent $i$ and drop the corresponding subindex in ${\mathcal L}_i$ and $U_i$ just denoting the space as ${\mathcal L}$ and the utility function $U$. We assume that the space of options $\mathcal{L}$ is a (real) Hilbert space. That is, it is a complete metric space with an inner product.  On the other hand, to ensure the existence of a maximal $\hat{{\mathbf x}}$ we will further assume that $\hat{L}$ is a compact subset of ${\mathcal L}$, while $U$ is a continuous function. Let ${\mathbf x}^{*}$ be a solution to the problem of optimizing $U$ over $\hat{L}$.

Consider now a family $\{L^{k}\}_{k=0}^{\kappa}$ of linear subspaces of ${\mathcal L}$, each one the domain of a {\em local} problem. For each $k = 0, \ldots, \kappa$ we define a projection: 

 \[\mbox{Proj}_k: {\mathcal L} \rightarrow \bigcup_{k=0}^{\kappa} L^{k}\] 
 
$\mbox{Proj}_k(x) = x^{k} \in L^{k}$, where $x^k$ is the {\em projection} of $x \in \mathcal{L}$ on $L^{k}$.\footnote{According to the Linear Projection Theorem,  $|x - x^k|= \min_{y \in L^{k}} |x -y|$, where $|\cdot|$ is the norm of ${\mathcal L}$ \cite{Luenberger}.}

A compact subset $\hat{L}^k \subseteq L^k$ is the domain of a {\em local} maximization problem. The projection of a global solution ${\mathbf x}^{*}$ onto $\hat{L}^k$ will return the points that are the closest to ${\mathbf x}^{*}$. If the projection does not return a local solution, it can be generalized to a map $\Gamma_k: \hat{L} \rightarrow \hat{L}^k$:

$$\Gamma_k(x) = \{x^k \in \hat{L}^{k} : x^k \in \mbox{argmin}_{y \in \hat{L}^{k}} |y - \mbox{Proj}_k(x)| \}.$$

That is, $\Gamma_k(x)$ yields points in $\hat{L}^k$ that are the closest to the orthogonal projection of $x$ on the subspace $L^k$ of $\mathcal{L}$. Then, given any solution $\hat{{\mathbf x}}^{k}$ of the local maximization problem over $\hat{L}^k$, we want to ensure that $\hat{{\mathbf x}}^{k} \in \Gamma_k(\mathbf{x}^{k})$.

More importantly, if the global solution is not given, we seek to find it by gluing together local ones, using the inverse of $\Gamma_k$. To formalize this we will introduce a category of local problems. This means that, given a family of local solutions $\{ \hat{{\mathbf x}}^{k} \}_{k=0}^{\kappa}$, we seek a ${\mathbf x} \in \mathcal{L}$ such that $\Gamma_k(\mathbf{x}) = \hat{{\mathbf x}}^{k}$ for each $k$.

We will start solving this problem by providing a sheaf-theoretic characterization of the abduction.

\section{The global solution: a sheaf-theoretic result}

To obtain $\hat{{\mathbf x}}$ (the ``global'' solution) by patching up the ``local'' maxima, $\{\hat{{\mathbf x}}^{k}\}_{k=0}^{\kappa}$ we resort to some previous definitions:\footnote{In this section we draw heavily on the literature on category theory, even if the results are elementary. Although almost self-contained, this analysis uses notions that can be found in \cite{Goldblatt}, \cite{Barr}, and \cite{Adamek}.}

\begin{defi}~\label{defiPR}
Let ${\mathbf {\mathcal PR}}$ to be the category of local problems, where
\begin{itemize}
\item The objects are local problems $s^k$ of the form $\langle \hat{L}^{k}, u^{k}, \hat{{\mathbf X}}^{k} \rangle$ where a continuous {\em local} utility function $u^{k}$ is maximized over a compact set $\hat{L}^{k} \subset L^{k}$, a linear subspace of ${\mathcal L}$, yielding a family of solutions $\hat{{\mathbf X}}^{k}$.
\item A morphism $\rho_{kj}: s^{k} \rightarrow s^{j}$ is defined as $\hat{L}^k \subseteq \hat{L}^j$, $u^k = u^{j}|_{L^k}$ and $\mbox{dim}(L^k) \leq \mbox{dim}(L^j)$.\footnote{$\mbox{dim}(\cdot)$ yields the dimension of a subspace of ${\mathcal L}$.} It follows from this definition that an identity morphism $\rho_{kk}: s^k \rightarrow s^k$ trivially exists for every object $s^k$. Furthermore, given two morphisms $\rho_{kj}: s^k \rightarrow s^j$ and $\rho_{jl}: s^j \rightarrow s^l$ there exists their composition $\rho_{jl}\circ \rho_{kl} = \rho_{kl}$, since $\hat{L}^k \subseteq \hat{L}^j \subseteq \hat{L}^l$, $ \mbox{dim}(L^k) \leq  \mbox{dim}(L^j) \leq  \mbox{dim}(L^l)$ and by transitivity of the restrictions $u^k = u^{j}|_{L^k}$ and $u^j = u^{l}|_{L^j}$ we have $u^k = u^{l}|_{L^k}$.
\item Given two local problems $s^k$ and $s^j$, if $\hat{L}^k \cap \hat{L}^j \neq \emptyset$, a problem $s^{k \wedge j}$ with that domain has, by the properties of morphisms, has to verify that $u^{k \wedge j}$ is such that
\[ u^{k}|_{L^k \cap L^j} \ = \ u^{k \wedge j} \ = \ u^{j}|_{L^k \cap L^j}\]
\end{itemize}
\end{defi}

We can also define ${\mathcal P}({\mathcal L})$ as the category in which the objects are subsets of ${\mathcal L}$ and a morphism between two objects $f_{AB}: A \rightarrow B$ is defined as $A \subseteq B$.

Now we define a functor $F: {\mathbf {\mathcal PR}}^{\mbox{op}} \rightarrow {\mathcal P}({\mathcal L})$, where ${\mathbf {\mathcal PR}}^{\mbox{op}}$ is the dual category of ${\mathbf {\mathcal PR}}$.\footnote{The objects of ${\mathbf {\mathcal PR}}^{\mbox{op}}$ are the same of ${\mathbf {\mathcal PR}}$, while for each morphism $\rho_{kj}: s^k \rightarrow s^j$ in ${\mathbf {\mathcal PR}}$ there is a $\rho_{kj}^{\mbox{op}}: s^j \rightarrow s^k$ in ${\mathbf {\mathcal PR}}^{\mbox{op}}$.}.

\begin{defi}
$F: {\mathbf {\mathcal PR}}^{\mbox{op}} \rightarrow {\mathcal P}({\mathcal L})$ is such that:
\begin{itemize}
\item For each object $s^k$ in ${\mathbf {\mathcal PR}}^{\mbox{op}}$, $F(s^k)= \{{\mathbf x} \in {\mathcal L}: \mathbf{x} \in \Gamma_k^{-1}( \hat{{\mathbf X}}^{k}) \}$.
\item $F(\rho_{kj})$ is defined for any morphism $\rho_{kj}^{\mbox{op}}: s^j \rightarrow s^k$ in ${\mathbf {\mathcal PR}}^{\mbox{op}}$ (that is, $\rho_{kj}: s^k \rightarrow s^j$ in ${\mathbf {\mathcal PR}}$) as the inclusion $F(s^j) \subseteq F(s^k)$. It follows trivially that $F$ preserves identities and compositions of morphisms in ${\mathbf {\mathcal PR}}^{\mbox{op}}$.
\end{itemize}
\end{defi}

The following result ensures that $F$ is well-defined:

\begin{propi}~\label{morphism}
Given two local problems $s^k$ and $s^j$, such that there exists a morphism $\rho_{kj}: s^k \rightarrow s^j$ in ${\mathbf {\mathcal PR}}$, 
$$\{{\mathbf x} \in {\mathcal L}: \mathbf{x} \in \Gamma_j^{-1}( \hat{{\mathbf X}}^{j}) \} \ \subseteq \ \{{\mathbf x} \in {\mathcal L}: \mathbf{x} \in \Gamma_k^{-1}( \hat{{\mathbf X}}^{k}) \}.$$
\end{propi}
\noindent {\bf Proof}: {\it The existence of $\rho_{kj}: s^k \rightarrow s^j$ means, in particular, that $(1)$ $\hat{L}^k \subseteq \hat{L}^j$ and  $(2)$ $u^k = u^{j}|_{L^k}$.

By $(1)$, ${\mathcal L}\setminus \hat{L}^j \subseteq {\mathcal L}\setminus \hat{L}^k$. On the other hand, by $(2)$ $\hat{\mathbf{x}}^k$ optimizes $u^{j}|_{L^k}$ and $\hat{\mathbf{x}}^j$ maximizes $u^j$. Then, for any $\mathbf{x}$ such that $\Gamma_j(\mathbf{x}) = \hat{\mathbf{x}}_j$, it follows that $\Gamma_k(\mathbf{x}) = \hat{\mathbf{x}}_k$. Then $\{{\mathcal L}: \mathbf{x} \in \Gamma_j^{-1}( \hat{{\mathbf X}}^{j}) \}$$\subseteq$${\mathcal L}: \mathbf{x} \in \Gamma_k^{-1}( \hat{{\mathbf X}}^{k}) \}$.}

All this means that we have defined a {\em contravariant functor} $F: {\mathbf {\mathcal PR}} \rightarrow {\mathcal P}({\mathcal L}_i)$. In category-theoretical terms this means that $F$ is a ${\mathcal P}({\mathcal L}_i)$-valued {\em presheaf} on ${\mathbf {\mathcal PR}}$. If some additional properties are satisfied, $F$ would also be a {\em sheaf}. The importance of this is that it would mean that there exists an object $s^{*}$ in ${\mathbf {\mathcal PR}}$ such that $F(s^*)$ ``glues'' together all the objects $F(s^k)$ in ${\mathcal P}({\mathcal L})$. To see that this is indeed the case and understand exactly what this means, we need some previous considerations.

The category ${\mathcal P}({\mathcal L})$ includes products of families of objects. That is, given $A_1, A_2, \ldots, A_n$ objects in ${\mathcal P}({\mathcal L})$, it exists an object $\prod_j A_j$ such that given projections $p_j: \prod_j A_j \rightarrow A_j$, if there exists another object $C$ and morphisms $f_j: C \rightarrow A_j$ there exists a single $p: C \rightarrow \prod_j A_j$ such that $p_j\circ p = f_j$ for every $j$. But recall that in ${\mathcal P}({\mathcal L})$ morphisms are inclusions of sets. Then $\prod_j A_j$ is the largest subset common to every $A_j$, that is, $\prod_j A_j = \bigcap_j A_j$.

In ${\mathbf {\mathcal PR}}$, the problem $s^{j \wedge l}$ is, as indicated in Definition~\ref{defiPR}, interpreted as the local problem in which the set of alternatives is $\hat{L}^{j} \cap \hat{L}^{l}$. The utility function $u^{j \wedge l}$ is the restriction of the continuous functions $u^{j}$ and $u^{l}$ over this domain, and is thus also continuous. The corresponding morphisms $\rho_{\{j,l\} \ j}:s^{j \wedge l} \rightarrow s^j$ and $\rho_{\{j,l\} \ j}:s^{j \wedge l} \rightarrow s^j$, which in ${\mathcal P}({\mathcal L})$ induce, in turn, the morphisms $c_{j,l}:F(s^j) \rightarrow F(s^{j \wedge l})$ and $d_{j,l}: F(s^l) \rightarrow F(s^{j \wedge l})$. Varying $j$ and $l$ we obtain two morphisms $c$ and $d$ such that:
$$ \prod_k F(s^k) \stackrel{c}{\rightarrow} \prod_{j,l} F(s^{j \wedge l}) \ \ \ \mbox{and} \ \ \ \prod_k F(s^k) \stackrel{d}{\rightarrow} \prod_{j,l} F(s^{j \wedge l})$$

Finally, in ${\mathbf {\mathcal PR}}$ an object $s^{*} = \max_{k}(s^k)$ can be defined, in which the set of alternatives is $\hat{L}^{*} = \bigcup_k \hat{L}^{k}$ and $u^{*}$ which is a continuous function that satisfies that for each compact  $\hat{L}^{k^{'}} \subseteq \hat{L}^{*}$ corresponding to a problem $s^{k^{'}}$, $u^{k^{'}} = u^{*}|_{\hat{L}^{k^{'}}}$. It is a well-defined object of ${\mathbf {\mathcal PR}}$ that is such that for every $k$ there exists a morphism $\rho_{* \ k}: s^k \rightarrow s^{*}$ and a corresponding $\epsilon_k \equiv F(s^k, s^{*})$$: F(s^{*}) \rightarrow F(s^{k})$. The latter induces in ${\mathcal P}({\mathcal L}_i)$ the morphism $\epsilon: F(s^{*}) \rightarrow \prod_k F(s^k)$.

Then, consider the following definition:

\begin{defi}[Barr \& Wells 1999]
$F: {\mathbf {\mathcal PR}}^{\mbox{op}} \rightarrow {\mathcal P}({\mathcal L})$ is a sheaf if given $s^{*}$ in ${\mathbf {\mathcal PR}}$, $\epsilon$ is the {\em equalizer} of $c$ and $d$.
\end{defi}

We have:

\begin{propi}~\label{equalizer}
$F: {\mathbf {\mathcal PR}}^{\mbox{op}} \rightarrow {\mathcal P}({\mathcal L})$ is a sheaf.
\end{propi}
\noindent {\bf Proof}: {\it It suffices to show that $\epsilon$ is an equalizer of $c$ and $d$. This means that $(1))$ $c\circ \epsilon = d\circ \epsilon$ and $(2)$ for any other $f: A \rightarrow \prod_k F(s^k)$ for a ${\mathcal P}({\mathcal L})$ object $A$, such that $c \circ f = d \circ f$ there exists a unique $g: A \rightarrow F(s^{*})$ verifying that $\epsilon \circ g = f$. But in ${\mathcal P}({\mathcal L})$ the morphisms are set inclusions. By the characterization of products, it means that $F(s^{*}) \subseteq \bigcap_k F(s^k) \subseteq \bigcap_{j,l} F(s^{j \wedge l})$, satisfying $(1)$.

Suppose that $A \subseteq \bigcap_k F(s^k) \subseteq \bigcap_{j,l} F(s^{j \wedge l})$, such that $A \not\subseteq F(s^*)$. This means that there exists an option ${\mathbf x} \in A \subseteq \bigcap_k F(s^k)$ such that ${\mathbf x} \notin F(s^{*})$. This means that ${\mathbf x} \in \Gamma_k^{-1}( \hat{{\mathbf X}}^{k})$ for every $k$ but $\Gamma_k ({\mathbf x})$ does not maximize $\max_k(u^k)$ over $\bigcup_k \hat{L}^{k}$. But then, $\Gamma_k({\mathbf x})$ maximizes {\em each} $u^k$ when restricted to $\hat{L}^{k}$. Since for every compact set $\hat{L}^{k^*}$, $u^{k} = u^{*}|_{\hat{L}^{k}}$, we have that $u^{*}[\Gamma_k({\mathbf x})] = u^k[\Gamma_k({\mathbf x})]$. Contradiction. Then $(2)$ is proven, showing that $\epsilon$ is the equalizer of $c$ and $d$.} 

\begin{exi} Consider ${\mathcal L} = \mathbb{R}^2$ and two problems:
\begin{itemize}
\item $s^k$: $\hat{L}^k = [0, 1] \times \{0\}$ and $u^k: x$. Thus ${\mathbf X}^k = \{ (1,0)\}$.
\item $s^j$: $\hat{L}^k = \{(x,y): 0 \leq x + y \leq 1\} $ and $u^k: x + 2y$. Thus ${\mathbf X}^k = \{ (0,1)\}$.
\end{itemize}

We can see that $\mbox{min}(s^k, s^j) = s^k$ and since $F(s^j) \subseteq F(s^k)$, a straightforward application of Proposition~\ref{equalizer} we have that $s^* = s^j$ and thus $s^*$ is such that $\hat{{\mathbf X}}^* = \{(0,1)\}$ with $\Gamma_j[(0,1)] = (0,1)$ and $\Gamma_k[(0,1)] = (1,0)$.
\end{exi}

The importance of Proposition~\ref{equalizer} is that any ${\mathbf x} \in  \hat{{\mathbf X}}^{*}$, when restricted on each $\hat{L}^{k}$ yields an element $\Gamma_k({\mathbf x}) \in \hat{{\mathbf X}}^{k}$. That is, up from local solutions we have obtained a global one. To see the features of such global solutions we can consider two cases: one when each $u_i^k$ is concave and a general one, which may be obtained via polynomial approximations.

\section{Concave utility functions}
Following the tradition in Economics we consider the case of a concave $\max_k(u^k)$, which by simplicity we will identify with $U$. Furthermore, $\hat{L} = \bigcup_k \hat{L}^{k}$. This means that each $u^{k}$ is the continuous and concave restriction of $U$ over $L^{k}$. It follows that:

\begin{propi}~\label{unique0}
If $U$ is strictly concave and $u^{k}$$\equiv$$U|_{L^{k}}$ then $\hat{{\mathbf x}}^{k}$$=$$\Gamma_k(\hat{{\mathbf x}})$.
\end{propi}
\noindent {\bf Proof}: {\it Suppose that $\hat{{\mathbf x}}^{k}$$\neq$$\Gamma(\hat{{\mathbf x}})$. Then,
 $$\mbox{argmax}_{x \in \hat{L}^{k}} u^{k}(x) \neq \mbox{argmax}_{x \in \hat{L}^{k}}U|_{L^{k}}(x).$$
But this is absurd, since $u^{k}$$\equiv$$U|_{L^{k}}$, $\hat{L}^{k} = \hat{L}|_{L^{k}}$, and $U$ and $u^k$, being continuous and strictly concave, yield {\em unique} solutions.}   

Then, the solution of a sequence of problems $\{s^{k}\}_{k=0}^{\kappa}$, yields a family $\{\hat{{\mathbf x}}^{k}\}_{k=0}^{\kappa}$, such that for each $s^{k}$, $\hat{{\mathbf x}}^{k}$$=$$\Gamma(\hat{{\mathbf x}})$.


Then, we have that:

\begin{propi}~\label{unique} Assume that $\{s^{k}\}_{k=0}^{\kappa}$ is a family of problems in which each $u^k$ is the continuous and strictly concave restriction of a continuous and strictly concave function $U$ over a compact $\hat{L} \subseteq \mathcal{L}$. If $\{d_k\}_{k=0}^{\kappa}$ is the set such that for each $k$,  $d_k = dim(L_k)$ and the solutions $\{{\mathbf x}^{k}\}_{k=0}^{\kappa}$ are linearly independent and span $\mathcal{L}$, a unique approximation $\tilde{\mathbf{x}}$ to the global maximum $\hat{{\mathbf x}}$ can be uniquely determined if $\sum_{k=0}^{\kappa} d_k= dim(\mathcal{L})$.
\end{propi}
\noindent {\bf Proof}: {\it For each $\hat{{\mathbf x}}^{k}$, the unique solution of the maximization of the concave function $u^k$, consider $Proj^{-1}_k (\hat{{\mathbf x}}^k) \subseteq \Gamma^{-1}_k(\hat{{\mathbf x}}^k)$. For $k=0, \ldots, \kappa$, we have that $Proj^{-1}_k (\hat{{\mathbf x}}^k)$  yields an affine subspace of ${\mathcal L}$, say $l^{-1}_k$. Then, by the linear independence of $\{\hat{{\mathbf x}}^k\}_{k=0}^{\kappa}$, $\cap_{k=0}^{\kappa} l^{-1}_k $ is an affine subspace of $\mathcal{L}$. Since $\{\hat{{\mathbf x}}^{k}\}_{k=0}^{\kappa}$ is a basis of $\mathcal{L}$, $\hat{{\mathbf x}} \in \cap_{k=0}^{\kappa} l^{-1}_k \neq \emptyset$ and since $\sum_{k=0}^{\kappa} = dim(\mathcal{L})$, there exists at most one point in that intersection, which we denote as $\tilde{{\mathbf x}}$. Notice that, in general, $\tilde{{\mathbf x}} \neq \hat{{\mathbf x}}$. But, by construction, $\tilde{}{{\mathbf x}}$ is the unique element in $\mathcal{L}$ such that, for each $k$ minimizes $|Proj_k(\tilde{\mathbf{x}}) - Proj_k (\hat{\mathbf{x}})|$.}

While the conditions of linear independence of local maxima and the dimensionality of the ``covering'' of the global problem by the local ones may seem too demanding, they indicate that, to recover the global maximum, the family of local problems should be {\em informative}. That is, we can approximate the global solution by putting together the scattered pieces of information provided by the local solutions.

As it is well known, a sufficient condition for $\tilde{\mathbf{x}} = \hat{\mathbf{x}}$ is that each $s^k$ must be such that:

\begin{itemize}
\item $\hat{L}$ can be written as $\hat{L} = \hat{L}^k \times \hat{L}^{k \perp}$, where $\hat{L}^{k \perp}$ is a compact set in which its elements are orthogonal to those in $\hat{L}^{k}$.
\item $0 \in \hat{L}^{k \perp}$, for $0 \in {L}^{k \perp}$, the subspace of $\mathcal{L}$ of which $\hat{L}^{k \perp}$ is a subset.
\item $U = u^k + u^{k \perp}$, such that $U(\hat{\mathbf{x}}) = u^k(\hat{\mathbf{x}}^{k}) + u^{k \perp}(\hat{\mathbf{x}}^{k \perp})$, where $U, u^k$ and $u^{k \perp}$ are continuous strictly concave functions.
\end{itemize}

\begin{exi}
Consider $U(x,y) = 3 - x^2 - y^2 + 2y + x$, defined over the compact set $\{(x,y): 0 \leq x, y \leq 1\} \subset \mathbb{R}^2$. $U$ is concave. The global problem of the maximization of $U$ has a unique solution: $\hat{\mathbf{x}} = (\frac{1}{2}, 1)$. Then, consider two local problems:
\begin{itemize}
\item $s^1$: the maximization of the restriction of $U$ over the compact segment $[0, 1]$ of the $1$-dimensional subspace consisting of the $x$-axis of $\mathbb{R}^2$. The restriction is $u(x) = 3 - x^2 + x$, which yields the unique solution $\hat{\mathbf{x}}^1 = \frac{1}{2}$.
\item $s^2$: the maximization of the restriction of $U$ over the compact segment $[0, 1]$ of the $1$-dimensional subspace consisting of the $y$-axis of $\mathbb{R}^2$. The restriction is now $u(y) = 3 - y^2 + 2y$, which yields a solution $\hat{\mathbf{x}}^2 = 1$.
\end{itemize}
Since in $\mathbb{R}^2$, the solutions of the local problems are the linearly independent vectors $(\frac{1}{2},0)$ and $(0,1)$, and the sum of the dimensions of the local subspaces is $2$, the intersection $Proj_{1}(\frac{1}{2}) = \{(\frac{1}{2},y)\}$ and $Proj_{2}(1) = \{(x,1)\}$ yields the global solution $(\frac{1}{2},1)$.
\end{exi}

\section{Local-to-global via polynomial approximations}

Here we will consider the class of solutions to a family of problems $\{s^k\}_{k=0}^{m}$ such that $L^{*} \subseteq \cup_{k=0}^m L^k = {\mathcal L}$, is the domain of a problem $s^{*}$ with utility $u^{*}$. We study the problem of finding a polynomial approximation to $u^{*}$, yielding a function $V$ with the same maxima. That is, given a family of problems in ${\mathbf {\mathcal PR}}$ we seek to find the simplest\footnote{By {\em simplest} we mean that it can be defined piecewise by polynomials of the lowest possible degree} continuous function $V: {\mathcal L} \rightarrow \mathbb{R}$ such that the $\Gamma$-projections of its maxima in $L^*$ are in $\hat{{\mathbf X}}^{k}$, for $k \in \{0, \ldots, m \}$.\footnote{This relates our approach to the literature on {\em splines} \cite{Holland}.} 

The subspaces underlying $\{s^k\}_{k=0}^{m}$, $\{\langle L^{k} \rangle \}_{k=0}^{m}$ determine a partition of the entire space ${\mathcal L}$. The classes in the partition are of three types:
\begin{itemize}
\item Sets of the form $L^{k} \setminus \bigcup_{j \neq k; j \in \{0, \ldots, m\}}L^{j}$. That is, the subsets of the local spaces that do not overlap with any other of the given subspaces.
\item Sets of the form $\bigcap_{k \in \gamma \subseteq \{0,\ldots, m\}} L^{k}$, such that there is no $\gamma^{\prime} \subseteq \{0,\ldots, m\}$ with $\bigcap_{j \in \gamma^{\prime}} L^{j}$$\subseteq$$\bigcap_{k \in \gamma} L^{k}$. This means that each intersection among subspaces is decomposed in minimal intersections, each one being a class in the partition.
\item ${\mathcal L} \setminus \bigcup_{k=0}^{m}L^{k}$. That is, the entire space without the subspaces corresponding to the local problems $s^0, \ldots, s^m$.
\end{itemize}

Notice that sets in these classes are not necessarily compact nor convex. Nevertheless we can define $\mathbf{F}_{m}^{r}$ to be the class of continuous piecewise multidimensional polynomials with real coefficients of degree at most $r$, such that each $f \in \mathbf{F}_{m}^{r}$ is $f:{\mathcal L} \rightarrow \mathbb{R}$, defined on the aforementioned partition of ${\mathcal L}$:

$$
f(x) \ = \ \cases{p^k(x), & if $x \in L^{k}\setminus \bigcup_{j \neq k}L^{j}$ \cr 
\ p^{\gamma}(x), & if $x \in \bigcap_{k \in \gamma} L^{k}$, and $x \notin \bigcap_{j \in \gamma^{\prime}} L^{k}$, for $\gamma^{\prime} \neq \gamma$\cr
p^{\alpha}(x), & if $x \in {\mathcal L} \setminus \bigcup_{k=0}^{m} L^{k}$. \cr}
$$

\noindent where each $p^{k}(x)$ or $p^{\gamma}(x)$ is a polynomial defined on $x_{1}, \ldots, x_{\lambda}$, such that $x = \sum_{j =1}^{\lambda} x_j e_j$ where $\{e_1, \ldots, e_{\lambda}\}$ is the orthonormal basis of ${\mathcal L}$. They have the lowest degree such that $\hat{{\mathbf x}}^{k}$, expressed in terms of the $e_{j}$s is a maximum of $u^k$ over $\hat{L}^k \subseteq L^{k}$ (if it does not belong to any other subspace $L^{j}$) or over $\bigcap_{k \in \gamma} L^{k}$. The polynomial $p^{\alpha}$, instead, describes the value of $f$ over all the possible local problems not yet tested. 

If ${\mathcal B}(S)$ is the {\em boundary} of a subset $S \subseteq {\mathcal L}$, the continuity of each $f$ is ensured by requiring that:\footnote{From now on, we assume that $\gamma$ yields a minimal intersection among subspaces.}
\begin{itemize}
\item if $x \in {\mathcal B}(L^{k}\setminus \bigcup_{j \neq k}L^{j}) \cap {\mathcal B}(\bigcap_{k \in \gamma} L^{k})$, then $p^{k}(x) = p^{\gamma}(x)$,
\item if $x \in {\mathcal B}(L^{k}) \cap {\mathcal B}({\mathcal L}\setminus \bigcup_{k=0}^{m} L^{k})$, then $p^{k}(x) = p^{\alpha}(x)$,
\item if $x \in {\mathcal B}(\bigcap_{k \in \gamma} L^{k}) \cap {\mathcal B}({\mathcal L} \setminus \bigcup_{k=0}^{m} L^{k})$, then $p^{\gamma}(x) = p^{\alpha}(x)$.
\end{itemize}

The conditions on the boundaries ensure that for the characterization of $f$, each class in the partition of ${\mathcal L}_i$ is compact. This is particularly important for the existence of $f$:

\begin{lemi}
$f$ is well-defined.
\end{lemi}
\noindent {\bf Proof}: {\it It is easy to see that $p^k$ and $p^{\gamma}$, satisfying the boundary conditions, are well-defined (Propositions~\ref{unique0} and \ref{unique} elaborate this point). The definition of $p^{\alpha}$ is more involved. First, let us focus on the components of $\bigcup_{k=0}^{m} L_{i}^{k}$. Nothing ensures that for any pair $x, x^{\prime} \in \bigcup_{k=0}^{m} L_{i}^{k}$ there exists a curve $[x, x^{\prime}]$, joining them, entirely contained in $\bigcup_{k=0}^{m} L_{i}^{k}$. If such curve exists, we take ${\mathcal B}(\bigcup_{k =0}^{m} L_{i}^{k}) \cap {\mathcal B}({\mathcal L}_i \setminus \bigcup_{k=0}^{m} L_{i}^{k})$, which by definition admits a continuous function (the polynomials of the forms $p^k$ and $p^{\gamma}$ on it). Then, a straightforward application of the {\em Brouwer-Urysohn-Tietze Extension Lemma} (\cite{Ok}) allows to find a continuous function over the entire space, coinciding with the continuous function on the boundary (we just need to consider the constraint on ${\mathcal L}_i \setminus \bigcup_{k=0}^{m} L_{i}^{k}$). Then, the {\em Stone-Weierstrass Approximation Theorem} ensures the existence of $p^{\alpha}$ arbitrarily close to the extended continuous function (\cite{Prenter}).}

{\it By a direct application of the {\em Hahn-Banach Theorem} we can find hyperplanes in ${\mathcal L} \setminus \bigcup_{k=0}^{m} L^{k}$ pairwise separating the components. The intersection of those hyperplanes defines (not necessarily ``rectangular'') boxes, each one containing a component. On each of these boxes, an application of the Brouwer-Urysohn-Tietze's Lemma yields a continuous extension. On any of the hyperplanes, say $H^{1,2}$, which separates boxes $B^1$ and $B^2$, let us consider a point $x$, and its value under the continuous function found for $B^1$, $f^{1}(x)$. Suppose that $f^{1}(x)\neq f^2(x)$, where $f^2$ is the continuous function on $B^2$. Then, we can build a continuous function $f$, such that $f(x)=f^1(x)$, for every $x \in B^1$. Since ${\mathcal L}$ separates points, we can find a hyperplane $H^{1,2 \prime}$ with the same normal vector as $H^{1,2}$, such that $f(x) = f^{2}(x)$ on $H^{1,2 \prime}\cap B^2$. For any $x \in B^2$, between $H^{1,2 \prime}$ and $H^{1,2}$, we can pick $x^1 \in H^{1,2}$ and $x^2 \in H^{1,2 \prime}$ such that $x \in [x^1, x^2]$, a segment collinear with the normal vector of both hyperplanes. Then define $f(x) = \frac{x - x^1}{x^2 - x^1} f(x^1) + \frac{x^2 - x}{x^2 - x^1} f(x^2)$. Then, $f$ becomes continuous over $B^1 \cup B^2$.}

{\it On lower-dimensional intersections between hyperplanes the same procedure can be applied, only now based on defining, instead of segments, higher-dimensional simplexes among points in the parallel hyperplanes and points in the intersections among the original hyperplanes. This yields again a continuous function. By means of this brute force approach, we obtain a continuous function over the entire ${\mathcal L} \setminus \bigcup_{k=0}^{m} L^{k}$. By the Stone-Weierstrass theorem, there exists a polynomial $p^{\alpha}$ arbitrarily close to this function.}

The properties of the polynomials $p^k$ and $p^\gamma$ are easy to obtain:

\begin{propi}
If $x \in L^{k}$ while $x \notin L^{j}$ for $j \neq k$, $f(x)$ is a polynomial of degree at most $2$.
\end{propi}
\noindent {\bf Proof}: {\it The degree of $p^{k}(x)$ is $2$ if $\hat{{\mathbf x}}^{k} \in \hat{L}^k$ is an interior point of $L^{k}$, since a quadratic expression ensures a single maximum. On the other hand, if $\hat{{\mathbf x}}^{k}$ is on ${\mathcal B}(L^{k})$, either a quadratic or a linear function (if the boundary is zero-dimensional) may yield it as a maximum.} 

On the other hand, the degree of $p^{\gamma}(x)$ depends on the number of local maxima in $\bigcap_{k \in \gamma} L^{k}$:\footnote{Here we lift the restriction of concavity of the utility functions.}

\begin{propi}~\label{quadratic}
\begin{enumerate}
\item If there is no $\hat{{\mathbf x}}^{k}$ in $\bigcap_{k \in \gamma} L^{k}$, $p^{\gamma}(x)$ is a constant function for every $x \in \bigcap_{k \in \gamma} L^{k}$. 
\item If there is a single $\hat{{\mathbf x}}^{k} \in \bigcap_{k \in \gamma} L^{k}$, $p^{\gamma}(x)$ is at most a quadratic function with a single maximum on $\hat{{\mathbf x}}^{k}$. 
\item If there are more than one local maxima in $\bigcap_{k \in \gamma} L^{k}$, the degree of $p^{\gamma}(x)$ is at least $3$.
\end{enumerate}
\end{propi}
\noindent {\bf Proof}: {\it Case $(1)$ follows from the fact that $p^{\gamma}(x)$ has the lowest possible degree, without having to reach a maximum on $\bigcap_{k \in \gamma} L^{k}$. A constant function satisfies these two conditions.}

{\it Case $(2)$ is identical to $p^{k}(x)$, since $\bigcap_{k \in \gamma} L^{k} \subseteq {\mathcal B}(L^{k})$.}

{\it Finally, case $(3)$ follows immediately from the fact that $p^{\gamma}(x)$ has more than one maximum and no polynomial function of degree less than $2$ yields a multiplicity of local maxima.} 

Then, it follows that:

\begin{teo}~\label{degree}
The highest degree $r$ of $p^{\gamma}(x)$ is equal to the highest number of local maxima in $\bigcap_{k \in \gamma} L^{k}$ for any $\gamma \subseteq \{0, \ldots, m\}$.
\end{teo}
\noindent {\bf Proof}: {\it Immediate from case $(3)$ in Proposition~\ref{quadratic} and the continuity of $f$.} 

These results do not yield a single $f$. Worse yet, the cardinality of $\mathbf{F}_{m}^{r}$ is infinite. Fortunately, we can drastically reduce the number of these alternatives. To see this, just consider $\mathbf{P}({\mathcal L})$, the set of polynomials that can be defined on ${\mathcal L}$. If the local solutions $\{\langle L^{k}, \hat{L}^k, \hat{{\mathbf x}}^{k}\rangle \}_{k=0}^{m}$ yield a partition of ${\mathcal L}$ in $\lambda$ classes, we have that $\mathbf{F}_{m}^{r}$ is a subset of $\mathbf{P}({\mathcal L})^{\lambda}$. On the other hand, if $\tau$ is the number of common boundaries between any two partitions,  $\mathbf{P}({\mathcal L})^{\tau}$ denotes the class of vectors in which the components are polynomials defined over the boundaries. Both $\mathbf{P}({\mathcal L})^{\lambda}$ and $\mathbf{P}({\mathcal L})^{\tau}$ are {\em modules}, i.e. generalizations of vector spaces in which the scalars are now elements of the {\em ring} of polynomials $\mathbf{P}({\mathcal L})$ (\cite{Romer}).

We can define a linear transformation $T: \mathbf{P}({\mathcal L})^{\lambda} \rightarrow \mathbf{P}({\mathcal L})^{\tau}$, where $T$, being linear, can be given a matrix representation with:
\begin{itemize}
\item $\tau$ rows,
\item $\lambda$ columns,
\item each element $jl$ in the matrix is $0, 1$ or $-1$ indicating if $\tau_j$ is a boundary of $\lambda_l$ or not,
\item each row $\tau_j$ contains only two non-zero entries, $1$ in $jl$ and the other $-1$, in $jk$, corresponding to the two partitions $\lambda_l$ and $\lambda_k$ that share the boundary $\tau_j$.
\end{itemize}

Then, for each $f \in \mathbf{F}_{m}^{r}$, $Tf$ yields an algebraic sum of the polynomials that constitute $f$, evaluated at the common boundaries. These sums can either be of the $p^k - p^{\gamma}$, $p^k - p^{\alpha}$ or $p^{\gamma} - p^{\alpha}$ types. By definition of $f$, these are all zero, and therefore $Tf = 0$. This means that $\mathbf{F}_{m}^{r}$ is the {\em kernel} of $T$.

It is a well known fact (see \cite{Mac Lane}) that the kernel of the transformation $T: \mathbf{P}({\mathcal L})^{\lambda} \rightarrow \mathbf{P}({\mathcal L})^{\tau}$ is a {\em submodule} of $\mathbf{P}({\mathcal L})^{\lambda}$. Then, Proposition $5.2.7$ in \cite{Cox} ensures the existence of a finite {\em Gr\"obner basis} of $\mathbf{F}_{m}^{r}$:\footnote{A {\em Gr\"obner basis} is a finite set of elements of the submodule that allows to express any element in it \cite{Sturmfels}.}

\begin{propi}
There exists $G = \{\bar{f}_{j}\}_{j \in J} \subset \mathbf{F}_{m}^{r}$, with $|J| < \infty$, such that any $f \in \mathbf{F}_{m}^{r}$, is $f \ = \ \sum_{j \in J} p^{j}\bar{f}_{j}$, where each $p^{j} \in \mathbf{P}({\mathcal L})^{\lambda}$.
\end{propi}

Then, given a family of local solutions $\{\langle L^{k}, \hat{{\mathbf x}}^{k}\rangle \}_{k=0}^{m}$, we can define a function $V^{m} \in \mathbf{F}_{m}^{r}$, $V^{m}(x) = \sum_{j \in J} \bar{f}_{j}$, where $\{\bar{f}_{j}\}_{j \in J}$ is the Gr\"obner basis of $\subset \mathbf{F}_{m}^{r}$.\footnote{This obtains by choosing $p^j(x) =1$ for every $x \in  {\mathcal L}$ and every $j \in J$.} In turn, the maximization over ${\mathcal L}$ of $V^{m}(x)$ yields a family $\{\hat{{\mathbf x}}^{*m}_j\}_{j=1}^{\mu}$, which we deem the maximal elements corresponding to $\{s^k \}_{k=0}^{m}$. Notice that $\mu \leq \lambda$, indicating that the number of maximal elements is at most equal to the number of classes in the partition of ${\mathcal L}$ given the local solutions.

\section{Evolving global solutions}

While $\hat{{\mathbf x}}^{*m}$ may be a close approximation of the intended global solution, further local solutions will allow its improvement. So, if an additional solution is included, namely a local subspace $L^{m+1}$, with a corresponding maximum $\hat{{\mathbf x}}^{m+1}$, a new family $\mathbf{F}_{m + 1}^{r^{\prime}}$ can be obtained.

\begin{propi}
$r^{\prime}$ is either $r$ or $r+1$.
\end{propi}
\noindent {\bf Proof}: {\it $r^{\prime}$, as seen in Theorem~\ref{degree}, equals the multiplicity of local maxima in intersections of the family $\{L^{k}\}_{k=0}^{m+1}$. Since $L^{m+1}$ may add only one more maximum to the intersections already obtained for $\{L^{k}\}_{k=0}^{m}$, then the number of local maxima is now either $r$ or $r+1$ (if the new maxima adds to the intersection where the multiplicity of maxima was exactly $r$).}

The particular refinement of the abduction performed for $m$ local problems is as follows:

\begin{teo}~\label{r}
$\mathbf{F}_{m + 1}^{r^{\prime}} \equiv \mathbf{F}_{m}^{r}$ over $\bigcup_{k=0}^{m} L^{k} \setminus L^{m+1}$.
\end{teo}
\noindent {\bf Proof}: {\it Suppose, by contradiction, that there exists $f^{\prime} \in \mathbf{F}_{m + 1}^{r^{\prime}}$ such that $f^{\prime} \neq f$ for every $f \in \mathbf{F}_{m}^{r}$. That is, for every $f \in \mathbf{F}_{m}^{r}$ there exists $x \in \bigcup_{k=0}^{m} L_{i}^{k}\setminus L^{m+1}$ such that $f^{\prime}(x) \neq f(x)$. But then, we have the following taxonomy of cases:
\begin{itemize}
\item $x$ is in a $L^{k}$ (without being in $L^{j}$ for $j \neq k$, $j \in \{0, \ldots, m+1\}$). By definition any $p^{\prime k}$ that is maximum over $\hat{{\mathbf x}}^{k}$ is acceptable, and there exists at least one $f \in \mathbf{F}_{m}^{r}$ such that $p^{k}(x) = p^{\prime k}$.
\item $x$ is in a $\bigcap_{k \in \gamma} L^{k}$ and $x \notin L^{m+1}$. Again, by definition any $p^{\prime \gamma}$ that ensures continuity with the $p^{\prime k}$s and maximality (if there is a maximum in the intersection) is acceptable and so, there is at least a $f \in \mathbf{F}_{m}^{r}$ such that $p^{\gamma}(x) = p^{\prime \gamma}$.
\end{itemize}
Therefore, there exists at least one $f \in \mathbf{F}_{m}^{r}$ such that $f^{\prime}(x) = f(x)$ for every $x \in \bigcup_{k=0}^{m} L^{k}\setminus L^{m+1}$. Contradiction.}

Of course, new functions may obtain, by defining polynomials of the lowest degree on $L^{m+1} \cap \bigcap_{k \in \gamma} L^{k}$ for $\gamma \subseteq \{0,\ldots,m\}$. The fact is that any new problem yields a finer partition of ${\mathcal L}$. On the classes of the new partition that were part of the original partition, the functions remain the same. But some classes become partitioned in finer classes. The only proviso is that the old polynomials should coincide in the new boundaries with the new ones.

By finding the Gr\"obner basis of $\mathbf{F}_{m + 1}^{r^{\prime}}$, we obtain now $V^{m+1}(x)$ and $\{\hat{{\mathbf x}}^{*m +1}_j\}_{j=1}^{\mu(m+1)}$. The question we want to explore now is whether any of the $\hat{{\mathbf x}}^{*(m+1)}_j$ comes close to a ``true'' global maximum when $m$ becomes larger.

To make sense of this question we have to assume that we are back to the conditions in section 3, namely that there exists a global strictly concave utility function $U = u^{*}$ such that for every $k$, $u^{k}$$\equiv$$U|_{\hat{L}^{k}}$. If the local utility functions were not related to $U$, no relation between $V$ and $U$ could be predicated.

Then, we have that: 

\begin{teo}~\label{norm}
If ${\mathcal L} \subseteq \bigcup_{k =0}^{\kappa} L^{k}$, there exists ${\hat m} >0$ such that for $m > {\hat m}$ there exist maximal element $\hat{{\mathbf x}}^{*m}$ of $V^{m}(x)$ such that $|\hat{{\mathbf x}}^{*m} - \hat{{\mathbf x}}| = 0$, where $|\cdot |$ is the norm on ${\mathcal L}$.
\end{teo}
\noindent {\bf Proof}: {\it Since ${\mathcal L} \subseteq \bigcup_{k =0}^{\kappa} L^{k}$, consider a closed ball $B(\hat{{\mathbf x}}, \epsilon)$ of radius $\epsilon$, around $\hat{{\mathbf x}}$. $B(\hat{{\mathbf x}}, \epsilon)$ is compact. Then, since $B(\hat{{\mathbf x}}, \epsilon)$$\subseteq$$ \bigcup_{k =0}^{\kappa} L^{k}$, there exists a finite family $\{L^{k}\}_{k= k_1, \ldots, k_K}$ that covers $B(\hat{{\mathbf x}}, \epsilon)$. This means that $\hat{{\mathbf x}} \in \bigcup_{k=k_1}^{k_K} L^{k}$. By choosing these subspaces and the corresponding local solutions we obtain $\mathbf{F}_{k_K}^{r}$ and consequently $V^{k_K}$. 

Furthermore, by a straightforward application of Theorem~\ref{r}, we have that $V^{m}(\hat{{\mathbf x}}) = V^{k_K}(\hat{{\mathbf x}})$, for every $m \geq k_K$.
On the other hand, from $\hat{{\mathbf x}} \in \bigcup_{k=k_1}^{k_K} L^{k}$ we have that $\hat{{\mathbf x}}^{k_j} = \hat{{\mathbf x}}$ for some $j=1, \ldots, K$ (because of the fact that $u^{k}$$\equiv$$U|_{L^{k}}$). Then, one of the maximal elements of $V^{m}$ for $m > {\hat m}= k_K$ will be $\hat{{\mathbf x}}$.}

\section{Conclusions}

Theorem~\ref{norm} shows that even by groping in the dark, the global sheaf-theoretic solution could be found, by selecting problems that cover the whole space of possibilities. Of course, this is not always possible, since the local problems usually present themselves without concern for the global solution. But if the global $u^{*}$ function is concave, as it is usually assumed in most economic applications, then $V$ will yield the global maximum of $u^{*}$.

The results presented here are quite general, keeping to a minimum the requirements on functions and choice sets. From a Machine Learning perspective, this procedure could allow finding solutions without the risk of overfitting, exhibiting a behavior close to what is known for neural networks as {\em grokking} (\cite{Mohamadi}).

\noindent {\bf Data Availability Statement}: no data is linked to this paper.

\end{document}